\documentclass[fleqn,12pt,twoside]{article}

\usepackage{graphicx}
\usepackage[figuresright]{rotating}

\newcommand{\AmS}{{\protect\the\textfont2
  A\kern-.1667em\lower.5ex\hbox{M}\kern-.125emS}}

\usepackage{espcrc1}

\title{Measurements of Collins and Sivers asymmetries at COMPASS}

\author{Paolo Pagano\footnote{This paper will be published in the proceedings of the ``16$\mathrm{^{th}}$ International Spin Physics Symphosium SPIN 2004'', October 10-16, 2004, Trieste, Italy.} \address{Sezione INFN di Trieste\\
Padriciano 99, 34012 Padriciano (TS), Italy\\
E-mail: Paolo.Pagano@cern.ch } on behalf of the COMPASS collaboration}

\begin{document}
\maketitle

\begin{abstract}
COMPASS is a fixed target experiment presently running at CERN.
In 2002, 2003, and 2004 the experiment used a 160 GeV polarized muon beam coming from SPS and 
scattered off
a $^6$LiD target. The nucleons in the target can be polarized either 
longitudinally or transversely with respect to the muon beam and 20\% of the running time has been devoted to transverse 
polarization.
From the transverse polarization data collected in 2002, which correspond
to a total integrated luminosity of about 
200 pb$^{-1}$, the Collins and the Sivers asymmetries have been determined 
separately and the preliminary results are presented here. 

\end{abstract}

\section{The theoretical framework}

The cross-section for polarized deep inelastic scattering\cite{Barone:2001sp} of leptons off 
spin 1/2 hadrons can be expressed, at the leading twist, as a function of 
three independent quark distribution function: $q (x)$, $\Delta q (x)$ and 
$\Delta_T q (x)$. The latter is chiral-odd and can be
measured in combination with a chiral-odd fragmentation function, the 
Collins function $\Delta D_a^h (z, p^h_T)$,
via azimuthal single spin asymmetries \cite{Collins:1992kk} (SSA) in the hadronic end-product 
(semi-inclusive measurement).
A similar effect can arise from a possible quark $k_T$ structure of a 
transversely polarized nucleon (the Sivers function, $\Delta^T_0 q$), which also causes
an azimuthal asymmetry in the produced hadrons.
Leptoproduction on transversely polarized nucleons is a favourable setting to disentangle the Collins and Sivers effects since they show a dependence from linearly independent kinematic variables.

%

According to Collins, 
the fragmentation function of a quark of flavour $a$ in a hadron $h$
can be written as\cite{Baum:1996yv}:
$$
D_a^h(z, {\bf p_T^{\, h}}) = D_a^h(z, p_T^{h}) + \Delta D_a^h(z, p_T^h) \cdot sin\Phi_C
$$
where
${\bf p_T^{\, h}}$ is the final hadron transverse momentum
with respect to the quark direction -- i.e. the virtual photon direction -- and
$z = E_h / (E_{l}-E_{l'})$ is the fraction of available energy 
carried by the hadron ($E_h$ is the hadron energy, $E_{l}$ is 
the incoming lepton
energy and $E_{l'}$ is the scattered lepton energy).
The angle appearing in the fragmentation function, known as ``Collins angle'' and noted as 
$\Phi_C$,
is conveniently defined in the system where the 
z-axis is the virtual photon direction and the x-z plane is the muon
scattering plane. 
In this frame 
$\Phi_C= \Phi_h -\Phi_s'$, where
$\Phi_h$ is the hadron azimuthal angle, and $\Phi_s'$ is the azimuthal angle
of the transverse spin of the struck quark. 
 Since $\Phi_s'=\pi-\Phi_s$, with $\Phi_s$ the azimuthal angle of the transverse spin of the initial quark (nucleon), the relation
$\Phi_C = \Phi_h +\Phi_s - \pi$ is also valid.
The fragmentation function $\Delta D_a^h(z, p_T^{\, h})$ couples to transverse spin distribution function $\Delta_T q (x)$ and gives rise to SSA (denoted as $A_{Coll}$) dependent on $x$, $z$ and $p_T^{\, h}$ kinematic variables.

Following the Sivers hypothesis, the difference in the probability of finding an unpolarised quark of transverse momentum $\bf k_T$ and $ - \bf k_T$ inside a polarised nucleon can be written as \cite{Anselmino2002}:
$$
{\mathrm P}_{q/p^\uparrow} (x, {\bf k_T}) - {\mathrm P}_{q/p^\uparrow} (x, - {\bf k_T}) =~
\sin\Phi_S~ \Delta^T_0 q ( x, k^2_T)
$$
where $\Phi_S= \Phi_k -\Phi_s$ is the azimuthal angle of the quark
with respect to the nucleon transverse spin orientation. It has been recently demonstrated by theoretical arguments \cite{Brodsky:2002,Collins:2002}, that SSA (denoted as $A_{Siv}$) coming from the coupling of the Sivers function with the un-polarised fragmentation function $D_a^h(z, p_T^{\, h})$ can be observed at the leading twist from polarised Semi-Inclusive DIS.

\section{The COMPASS 2002 run for transversity}

The COMPASS \cite{Baum:1996yv,Bressan:Spin2004} experiment makes use of a high energy, intense,
polarised muon beam naturally polarised by the $\pi -$ decay mechanism.
It uses the polarised target system of the SMC experiment,
which consists of two $^6$LiD cells, each 60 cm long,
located along the beam one after the other in two separate RF cavities.
Data are taken therefore on the two oppositely polarised target cells 
simultaneously. 

\begin{figure}[!ht] %
\begin{center}
\includegraphics[width=\textwidth]{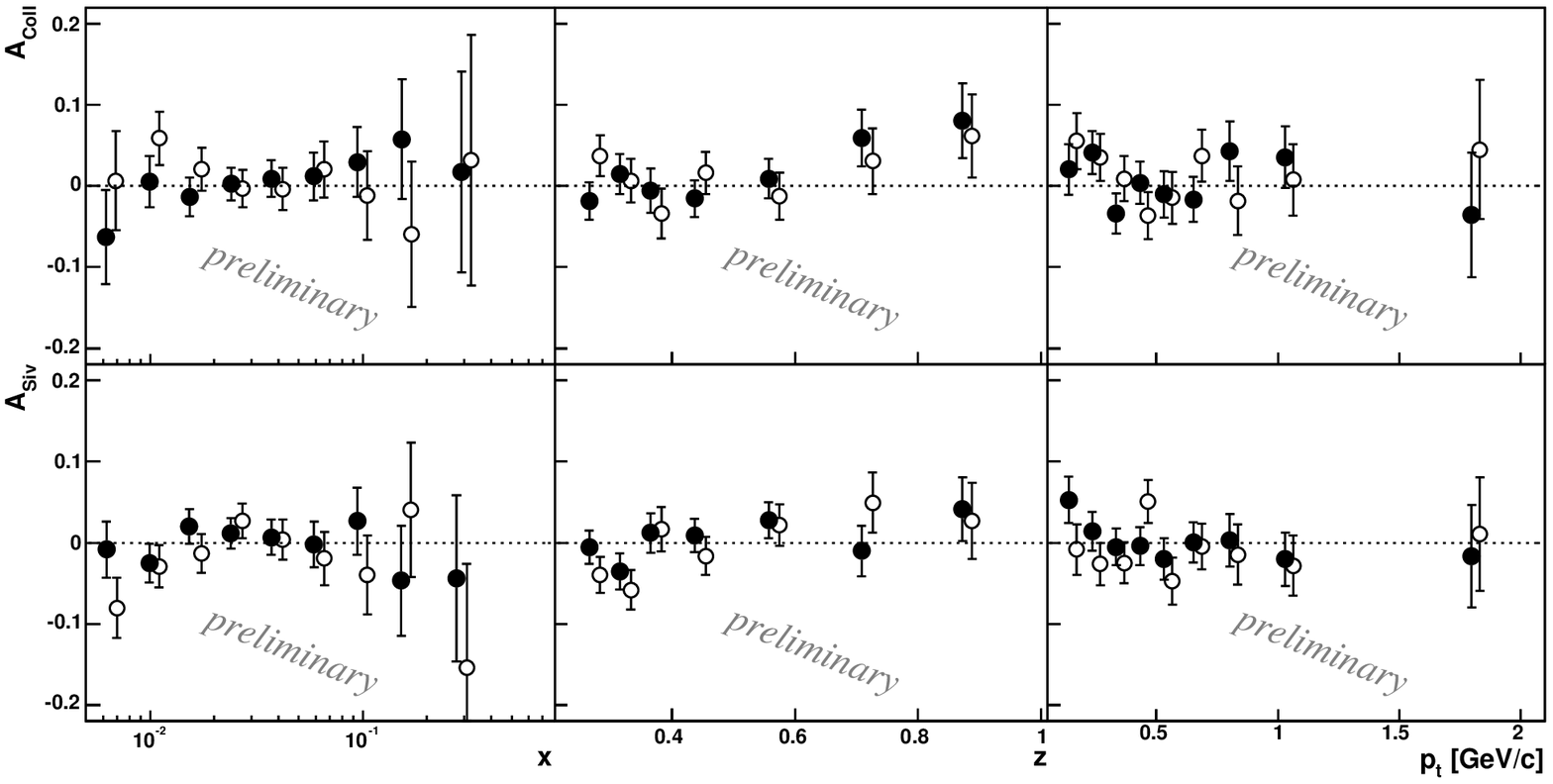}
\includegraphics[width=\textwidth]{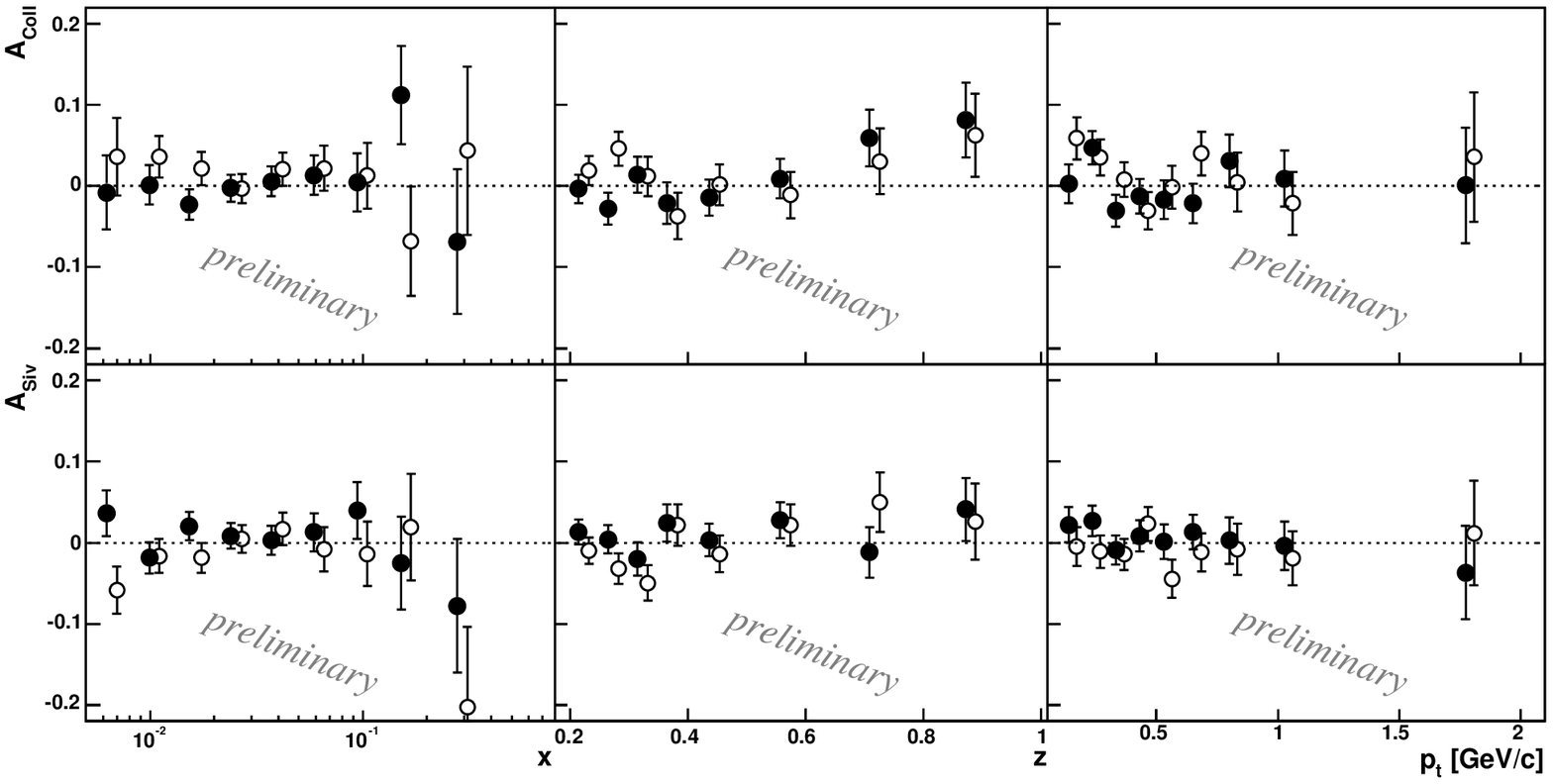}
\caption{Collins and Sivers asymmetry for positive (full points) and negative (open points) hadrons as a function of  $x$, $z$ and $p_T^{\, h}$. Leading hadrons analysis on top canvas, all hadrons on bottom canvas.}\label{results}
\end{center}
\end{figure}

Hereby we discuss the analysis of the data collected in year 2002 
with target polarisation oriented transversely to the beam direction.
This sample ( about 200 pb$^{-1}$ in integrated luminosity ) consists of two periods, each 5 days long, with two opposite settings
in the target spin orientation.
Events were selected in which a primary vertex (with identified beam and
scattered muon) was found in one of the two target cells with a least one
outgoing hadron. A clean
separation of muon and hadron samples was achieved by cuts on the amount of material traversed in the spectrometer. In addition, the kinematic cuts $Q^2 > 1$(G$e$V$/c)^2$, $W > 5$ G$e$V/$c^2$ and $0.1 < y < 0.9$ were
applied to the data to ensure a deep-inelastic sample above the region of
the nuclear resonances and within the COMPASS trigger acceptance. The upper
bound on $y$ also serves to keep radiative corrections small. SSA have been looked for both the leading hadron in the event, and for all the hadrons.
The leading hadron  was determined as the most energetic non-muonic
particle of the primary vertex having $z > 0.25$, and a transverse momentum
$p_T^{\, h} > 0.1$ G$e$V/$c$. 
When all the hadrons coming from the primary vertex were considered, the $z$ cut was lowered to $0.20$. The final data sample had an average values for $ x  = 0.034$, $ y  = 0.33$ and $ Q^2 =  2.7$ (G$e$V$/c)^2$. The average value for $z$ and $p_T^{\, h}$ are $0.44$ and $0.51$ G$e$V/$c$ for the leading hadron analysis, $0.38$ and $0.48$ G$e$V$/c$ in the other case.
In transverse polarisation, one can write the number of events as follows:
$$
N(\Phi_{C/S}) = \alpha(\Phi_{C/S}) \cdot N_0 \, (1 + \epsilon_{C/S} \sin{\Phi_{C/S}})\, ,
$$
\noindent where $\epsilon$ is the amplitude of the experimental asymmetry and $\alpha$
is a function containing the apparatus acceptance.
The former amplitude can be expressed as a function of the Collins and Sivers asymmetries through the expressions:
$$
 \epsilon_C = A_{Coll} \cdot P_T \cdot f \cdot D_{NN}  \qquad \mathrm{or} \qquad \epsilon_S = A_{Siv} \cdot P_T \cdot f\, ,\nonumber
$$
\noindent where $P_T$ ($\simeq 0.45$) is the polarisation of the target, $D_{NN}$ is the spin transfer coefficient, and $f$  ($\simeq 0.40$) is the target dilution factor.
To eliminate systematic effects due to acceptance,
in each period the asymmetry $\epsilon_{C}$ ($\epsilon_{S}$) is fitted separately for the two target cells from the event flux with
the two target orientations using the expression:
$$
\epsilon_{C/S} \sin \Phi_{C/S} = \frac{N^{\uparrow}_h(\Phi_{C/S}) - R \cdot N^{\downarrow}_h(\Phi_{C/S} + \pi)}
{N^{\uparrow}_h(\Phi_{C/S}) + R \cdot N^{\downarrow}_h(\Phi_{C/S} + \pi)}
$$
where $R = N^{\uparrow}_{h, tot}/N^{\downarrow}_{h, tot}$ is the ratio of the total number
of events in the two target polarisation orientations.
The results of the asymmetries plotted against the kinematic variables $x$, $z$ and $p_T^{\, h}$ are shown in Fig. \ref{results} for positive (full points) and negative (open points) hadrons, taking only the leading (top plot) hadron and for all hadrons (bottom plot). 

These are the first measurements of transverse spin effects on a deuteron target:
within the statistical accuracy of the data, both the Collins and Sivers asymmetries turned out to be small and compatible with zero, with a marginal indication of a Collins effect at large $z$ for both positive and negative charges.

\end{document}